\documentclass[12pt]{article}
\usepackage{graphicx}
\usepackage{a4wide}

\newcommand{\be}{\begin{equation}}
\newcommand{\ee}{\end{equation}}
\newcommand{\ba}{\begin{eqnarray}}
\newcommand{\ea}{\end{eqnarray}}
\newcommand{\trc}[1]{\mathrm{tr}_c\left(#1\right)}
\newcommand{\trf}[1]{\mathrm{tr}_F\left(#1\right)}
\newcommand{\one}{\mathrm{I}}

\begin{document}

\begin{titlepage}
\begin{flushright}
LU TP 09-27\\
October 2009
\end{flushright}
\vfill
\begin{center}
{\Large\bf Technicolor and other
QCD-like theories at next-to-next-to-leading order}
\vfill
{\bf Johan Bijnens and Jie Lu}\\[0.3cm]
{Department of Theoretical Physics, Lund University,\\
S\"olvegatan 14A, SE 223-62 Lund, Sweden}
\end{center}
\vfill
\begin{abstract}
We calculate the vacuum-expectation-value, the meson mass
and the meson decay constant to next-to-next-to-leading-order in
the chiral expansion for QCD-like theories with general $N_F$
degenerate flavours for the cases with a complex representation,
a real and a pseudoreal representation,
i.e. with global symmetry and breaking patters
$SU(N_F)_L\times SU(N_F)_R\to SU(N_F)_V$,
$SU(2N_F)\to SO(2N)$ and $SU(2N_F)\to Sp(2N_F)$. These calculations should
be useful for lattice calculations for dynamical electroweak
symmetry breaking and related cases.
\end{abstract}
\vfill
\end{titlepage}

\section{Introduction}
\label{introduction}

Chiral Perturbation Theory (ChPT) \cite{Weinberg1,GL1,GL2}
as effective field theory (EFT) for QCD is a very well
established method within strong interaction phenomenology.
The same method can also be used for different symmetry pattern cases.
These can be of interest for theories beyond the standard model.
Early papers in this context are the technicolor variations discussed in
\cite{Peskin,Preskill,Dimopoulos}.
Recently lattice calculations have started to explore some of these
cases, some recent references are \cite{lattice}.
While one is primarily interested in these theories in the massless
limit, lattice calculations are performed with finite masses
and the results thus need to extrapolated to zero mass.
For these extrapolations EFT is an excellent tool
and it is heavily used in fitting results for the pseudoscalar meson
octet in the QCD case. For high precision fits it is needed there to
go to next-to-next-to-leading-order in the ChPT expansion.

When writing the EFT relevant for dynamical electroweak symmetry breaking
one needs to consider different patterns of spontaneous breaking
of the global symmetry than in QCD. The resulting set of Goldstone Bosons,
or pseudo-Goldstone bosons in the presence of mass terms,
is thus also different. The low-energy EFT is thus also different.

In this paper we only discuss cases where the
underlying strong interaction is vectorlike and all
fermions have the same mass. Here three main patterns of global symmetry
show up. A thorough discussion of these cases at tree level
or lowest order (LO) is \cite{Kogut}.
With a gauge group with $N_F$ fermions in a complex representation
we have a global symmetry group $SU(N_F)_L\times SU(N_F)_R$
and we expect this to be spontaneously broken to the diagonal subgroup
$SU(N_F)_V$. This is the direct extension of the QCD case.
For $N_F$ fermions in a real representation the global symmetry group becomes
$SU(2N_F)$ and is expected to be spontaneously broken to $SO(2N_F)$.
In the case of two colours and $N_F$ fermions in the fundamental (pseudoreal)
representation the global symmetry group is again $SU(2N_F)$ but here
is expected to be spontaneously broken to an $Sp(2N_F)$ subgroup.
Some earlier references are \cite{Kogan,Leutwyler,SV}.
The complex case was treated to next-to-leading-order (NLO)
in \cite{GL2} in general and for the quantities considered here in \cite{GL4}.
The pseudo-real case has been done to NLO in \cite{Splittorff}.
We repeat here both calculations and also extend the third, real, case to
NLO by calculating the full infinity structure for all three cases
at NLO and giving the NLO Lagrangian.

In addition we also go to NNLO for three explicit quantities,
the vacuum-expectation-value, the meson mass and the meson decay constant
in the equal mass case. These formulas are our main result.
We expect that the NNLO Lagrangian for all cases will be a simple generalization
of the one for the complex case given in \cite{BCE1} but the calculation
of the general divergence structure, though in principle similar
to the one in \cite{BCE2}, we have not performed.

In the remainder of this paper we refer to the complex representation case
as QCD, the real representation case as adjoint and the
pseudo-real case as two-colour or $N_c=2$.
We first discuss in Sect.~\ref{quarklevel} the three different cases
at the underlying fermion (quark) level. Here we introduce explicit external
fields as done in \cite{GL1,GL2}. Sect.~\ref{EFT} introduces the
LO and NLO effective field theory and we do this using the
general formalism derived in \cite{CCWZ}. This allows to see
how similar the calculations for the three cases are.
In Sect.~\ref{infinities} we derive the divergent part at NLO
and in Sect.~\ref{NNLO} we calculate the NNLO result for the
meson mass, meson decay constant and vacuum expectation value.
In Sect.~\ref{conclusions} we summarize our results.

\section{Quark level}
\label{quarklevel}

This section shortly introduces the quark-level Lagrangian
giving the gauge groups and showing how the
condensates can be written in the more general cases.
A more extensive version of this discussion can be found
in \cite{Kogut} and the earlier references \cite{Kogan,SV}.
We remind the reader that we only consider cases with an underlying simple
vector gauge group  and we assume confinement and the formation of
a condensate.

We use the notation
$q_R$ and $q_L$ for the right- and left-handed fermions respectively.
Gauge indices we usually suppress and flavour indices will be indicated
when needed.

\subsection{QCD}
\label{quarkQCD}

This is the usual case where the fermions are in a complex representation
of the gauge group. With flavour indices $i$ the fermion part of the
Lagrangian enhanced by external fields is given by
\ba
\mathcal{L} &=&  \overline q_{Li}i\gamma^\mu D_\mu q_{Li} +
    \overline q_{Ri}i\gamma^\mu D_\mu q_{Ri}
   +\overline q_{Li}\gamma^\mu l_{\mu ij} q_{Lj}
   +\overline q_{Ri}\gamma^\mu r_{\mu ij} q_{Rj}
\nonumber\\&&
   -\overline q_{Ri}\mathcal{M}_{ij}q_{Lj}
   -\overline q_{Li}\mathcal{M}^\dagger_{ij}q_{Rj}\,.
\ea
The covariant derivative is given by $D_\mu q = \partial_\mu q-i G_\mu q$.

When the external fields vanish there is a $SU(N_F)_L\times SU(N_F)_R$
symmetry in the first two terms which is spontaneously
broken to the diagonal subgroup $SU(N_F)_V$.
This symmetry can be made local by adding the external fields
with the transformations $g_L\times g_R \in SU(N_F)_L\times SU(N_F)_R$:
\ba
\label{QCDtransformations}
q_L&\to& g_L q_L,\qquad q_R\to g_R q_R ,\qquad
 \mathcal{M}\to g_R\mathcal{M} g_L^\dagger\,,
\nonumber\\
l_\mu &\to& g_L l_\mu g_L^\dagger + i g_L\partial_\mu g_L^\dagger,\qquad
r_\mu\to g_R r_\mu g_L^\dagger + i g_R\partial_\mu g_R^\dagger\,.
\ea
We have here written $q_L$ and $q_R$ as column vectors in flavour and the
external fields $l_\mu$, $r_\mu$ and $\mathcal{M}=s-ip$ as matrices in flavour.

For later use, we define the big, $2N_F$, columnvector
\be
\hat q = \left(\begin{array}{c}q_R\\q_L\end{array}\right)
\ee
and the big, $2N_F\times 2N_F$, matrices
\be
\hat V_\mu =  \left(\begin{array}{cc} r_\mu & 0\\0 & l_\mu\end{array}\right)
\,\qquad
\hat \mathcal{M} =
\left(\begin{array}{cc} 0 & \mathcal{M}\\
\mathcal{M}^\dagger & 0\end{array}\right)\,,
\qquad\hat g =  \left(\begin{array}{cc} g_R & 0\\0 & g_L\end{array}\right)\,.
\ee
In terms of these the symmetry transformation can be written as
\be
\label{QCDtransformations2}
\hat q \to \hat g\hat q\,,\qquad \hat V_\mu
\to \hat g \hat V_\mu \hat g^\dagger+i \hat g \partial_\mu \hat g^\dagger
\,,\qquad
\hat\mathcal{M}\to \hat g \hat\mathcal{M}\hat g^\dagger\,.
\ee
Note that the symmetry group is not made larger since
$\hat q$ contains objects that have different Lorentz properties.

The formation of a flavour neutral condensate
$\langle \overline q q\rangle=\langle\overline q_R q_L\rangle+\mathrm{h.c.}$
breaks the full symmetry
spontaneously to the diagonal subgroup $SU(N_F)_V$

\subsection{Adjoint}
\label{quarkAdjoint}

If the fermions are in the adjoint representations we can write down
a similar Lagrangian as above
\ba
\label{lagrangianadjoint}
\mathcal{L} &=&  \trc{\overline q_{Li}i\gamma^\mu D_\mu q_{Li}} +
    \trc{\overline q_{Ri}i\gamma^\mu D_\mu q_{Ri}}
   +\trc{\overline q_{Li}\gamma^\mu l_{\mu ij} q_{Lj}}
   +\trc{\overline q_{Ri}\gamma^\mu r_{\mu ij} q_{Rj}}
\nonumber\\&&
   -\trc{\overline q_{Ri}\mathcal{M}_{ij}q_{Lj}}
   -\trc{\overline q_{Li}\mathcal{M}^\dagger_{ij}q_{Rj}}\,.
\ea
$\trc{A}$ means a trace over the gauge group indices and the fermions
are a matrix rather than a vector in the gauge group indices and
$D_\mu q = \partial_\mu q-iG_\mu q + i q G_\mu$.
Here we have the same transformation for the conjugated fermions,
$D_\mu\overline q = \partial_\mu-i G_\mu \overline q + i\overline q G_\mu$
with $\overline q = q^\dagger \gamma^0$ and the Hermitian conjugate also
means that the two gauge-indices are transposed.
The symmetries discussed here
exist in principle for any real representation for the fermions,
not only the adjoint one.

We define the matrix $C=i\gamma^2\gamma^0$ and we can define a new fermion
field
\be
\label{defqtilde}
\tilde q_{Ri} \equiv C \overline q_{Li}^T\,.
\ee
The transpose in (\ref{defqtilde}) works on the Dirac (and later also flavour)
indices but not on the gauge indices.
The field $\tilde q_{Ri}$ has the same transformation properties
under the gauge group as $q_R$ and is also a right-handed
fermion.\footnote{We have chosen right-handed rather than
left-handed in order to end up with transformations for fields
that look most like those for the QCD case in \cite{GL1,GL2}.}
In terms of the big matrices
\be
\hat q =\left(\begin{array}{c}q_R\\\tilde q_R\end{array}\right)\,,
\hat V_\mu =  \left(\begin{array}{cc} r_\mu & 0 \\
     0 & -l_\mu^T \end{array}\right)
\,\qquad
\hat \mathcal{M} =
\left(\begin{array}{cc} 0 & \mathcal{M}\\
\mathcal{M}^T & 0\end{array}\right)\,.
\ee
The Lagrangian (\ref{lagrangianadjoint}) becomes
\be
\label{lagrangianadjoint2}
\mathcal{L} =
    \trc{\overline {\hat q} i\gamma^\mu D_\mu \hat q}
   +\trc{\overline {\hat q}\gamma^\mu \hat V_\mu \hat q_{j}}
   -\frac{1}{2}\trc{\overline{\hat q} C \hat \mathcal{M}\overline{\hat q}^T}
   -\frac{1}{2}\trc{ \hat q^T C \hat\mathcal{M}^\dagger \hat q}\,.
\ee
The Lagrangian (\ref{lagrangianadjoint2})
has clearly a larger symmetry group, $SU(2N_F)$
as compared to QCD case above when we extend the external fields
to the full matrices and have as symmetry transformations:
\be
\hat q \to \hat g\hat q\,,\quad \hat V_\mu
\to \hat g \hat V_\mu \hat g^\dagger+i \hat g \partial_\mu \hat g^\dagger
\,,\quad
\hat\mathcal{M}\to \hat g \hat\mathcal{M}\hat g^T\,.
\ee
The Vafa-Witten argument shows that also in this case the vector symmetries
remain unbroken. We expect again a flavour neutral vacuum condensate
$\langle\trc{ \bar q q}\rangle$ which can be written as
$\langle\trc{ \hat q^T C J_S \hat q}\rangle+ \mathrm{h.c.}$
with
\be
J_S=
\left(\begin{array}{cc} 0 & \one\\
\one & 0\end{array}\right)
\ee
and $\one$ the $N_F\times N_F$ unit matrix.
This condensate breaks the the symmetry group down to $SO(2N_F)$.

\subsection{$N_c = 2$}
\label{quarkSU2}

The fundamental representation of $SU(2)$ is pseudo-real. The Lagrangian
enhanced with external fields reads
\ba
\label{lagrangianSU2}
\mathcal{L} &=&  \overline q_{Li}i\gamma^\mu D_\mu q_{Li} +
    \overline q_{Ri}i\gamma^\mu D_\mu q_{Ri}
   +\overline q_{Li}\gamma^\mu l_{\mu ij} q_{Lj}
   +\overline q_{Ri}\gamma^\mu r_{\mu ij} q_{Rj}
\nonumber\\&&
   -\overline q_{Ri}\mathcal{M}_{ij}q_{Lj}
   -\overline q_{Li}\mathcal{M}^\dagger_{ij}q_{Rj}\,.
\ea
The covariant derivative is given by $D_\mu q = \partial_\mu q-i G_\mu q$.

We can define a field $\tilde q_R$ as in the previous section via
\be
\tilde q_{R\alpha i} = \epsilon_{\alpha\beta} C \overline q_{L\beta i}^T\,,
\ee
with $\alpha,\beta$ gauge group indices,
$\epsilon_{12}=-\epsilon_{21}=1$, $\epsilon_{11}=\epsilon_{22}=0$
and $C=i\gamma^2\gamma^0$ as defined before.
The field $\tilde q_R$ is a right handed-handed fermion that transforms
as the fundamental representation of $SU(2)$.

In terms of the big matrices
\be
\hat q =\left(\begin{array}{c}q_R\\\tilde q_R\end{array}\right)\,,
\hat V_\mu =  \left(\begin{array}{cc} r_\mu & 0 \\
     0 & -l_\mu^T \end{array}\right)
\,\qquad
\hat \mathcal{M} =
\left(\begin{array}{cc} 0 & -\mathcal{M}\\
\mathcal{M}^T & 0\end{array}\right)\,.
\ee
The Lagrangian (\ref{lagrangianSU2}) becomes
\be
\mathcal{L} =
    \overline {\hat q} i\gamma^\mu D_\mu \hat q
   +\overline {\hat q}\gamma^\mu \hat V_\mu \hat q_{Lj}
   -\frac{1}{2} \overline{\hat q}_\alpha C\epsilon_{\alpha\beta}
 \hat \mathcal{M}\overline{\hat q}^T_\beta
   -\frac{1}{2} \hat q_\alpha\epsilon_{\alpha\beta}
   C \hat\mathcal{M}^\dagger \hat q_\beta\,.
\ee
This has again much larger symmetry group, $SU(2N_F)$
as compared to QCD case above when we extend the external fields
to the full matrices and have as symmetry transformations:
\be
\hat q \to \hat g\hat q\,,\quad \hat V_\mu
\to \hat g \hat V_\mu \hat g^\dagger+i \hat g \partial_\mu \hat g^\dagger
\,,\quad
\hat\mathcal{M}\to \hat g \hat\mathcal{M}\hat g^T\,.
\ee
The Vafa-Witten argument shows that also in this case the vector symmetries
remain unbroken and we expect again a flavour neutral vacuum condensate
$\langle \bar q q\rangle$ which can be written as
$\langle \hat q_\alpha\epsilon_{\alpha\beta} C J_A \hat q_\beta\rangle
+ \mathrm{h.c.}$
with
\be
J_A=
\left(\begin{array}{cc} 0 & -\one\\
\one & 0\end{array}\right)
\ee
and $\one$ the $N_F\times N_F$ unit matrix.
This condensate breaks the the symmetry group down to $Sp(2N_F)$.

\section{Effective field theory}
\label{EFT}

In this section we will show how the three cases can be brought into
an extremely similar form. That will allow to take over directly
much of the technology developed for the QCD case to the other cases.
We assume the reader to be familiar with ChPT and EFT.
Introductions can be found in \cite{lectures}.
We will use the terminology LO, NLO and NNLO for
the usual powercounting of order $p^2$, $p^4$ and $p^6$.

\subsection{QCD}
\label{EFTQCD}

The Goldstone bosons from the spontaneous symmetry breakdown
live in the space of possible vacua. For QCD and generalizations
this is in the form of a nonzero vacuum condensate
\be
\langle \overline q_{Lj} q_{Ri}\rangle =
\frac{1}{2}\langle \overline q q\rangle \delta_{ij}\,.
\ee
This vacuum is left unchanged by the vector transformations
with $g_L\times g_R\in SU(N_F)_L\times SU(N_F)_R$ and $g_L=g_R$.
The unbroken symmetry is $SU(N_F)$.
The broken symmetry part of the group are the axial transformations
wit $g_R=g_L^\dagger\equiv u$,
they rotate the vacuum into
\be
\langle \overline q_{Lj} q_{Ri}\rangle_\mathrm{rotated}
= \frac{1}{2}\langle \overline q q\rangle U_{ij}\,
\ee
with $U=g_R g_L^\dagger=u^2$.
The special unitary matrix $U$ describes the space of possible
vacua and varies under the symmetry as
\be
\label{Utransformation}
U\to g_R U g_L^\dagger\,.
\ee
This matrix $U$ can be used to construct the Lagrangians as was done
in \cite{GL2}. The covariant derivative on $U$ is defined as
\be
D_\mu U = \partial_\mu U -i r_\mu U +i U l_\mu\,.
\ee
The lowest order Lagrangian is
\be
\mathcal{L} = \frac{F^2}{4}\langle D_\mu U D^\mu U^\dagger
+ \chi U^\dagger + U \chi^\dagger\rangle\,,
\ee
with $\chi = 2 B_0 \mathcal{M}$ and $\langle A\rangle = \trf{A}$
This has the full global symmetry as can be checked using
the transformations (\ref{QCDtransformations}) and (\ref{Utransformation}).
In terms of the pion fields $\pi^a$ the matrix $u$ can be parametrized
as
\be
u = \exp\left(\frac{i}{\sqrt{2}\,F}\pi^a T^a\right)\,.
\ee
The $T^a$ are the generators of $SU(N_F)$ and normalized as $\trf{T^a T^b} =
\delta^{ab}$.

Let us now do the same analysis using the general formalism (CCWZ) \cite{CCWZ}.
We only look at the properties in the neighbourhood of the unit matrix
here. For the perturbative treatment we do here that is sufficient.
The global symmetry group $G$ has generators $T^a$
which are split up in a set of conserved generators $Q^a$ and broken
generators $X^a$. The $Q^a$ generate the unbroken symmetry group $H$
while the generators $X^a$ generate in a sense the manifold of
possible vacua, the quotient $G/H$ . We must now find a way to
parametrize the manifold $G/H$ and define covariant derivatives in general.
The manifold $G/H$ and the group $H$ we parametrize with
\be
\label{defuh}
\hat u = \exp\left(i\phi^a X^a\right) \in G/H\,,\qquad
\hat h = \exp\left(i\epsilon^a T^a\right) \in H\
\ee
The symmetry transformation we define using the property that
any group element $\hat g^\prime$ can be written in the form
\be
\label{defh}
\hat g^\prime = \hat u^\prime \hat h\,,
\ee
where both $\hat u^\prime$ and $\hat h$ are unique
and of the form (\ref{defuh}).
The symmetry transformation on $\hat u$ by a group element $\hat g\in G$
is defined as
\be
\hat u\to \hat g \hat u \hat h^\dagger
\ee
where $\hat h$ is the $\hat h$ of (\ref{defh}) needed to bring
$\hat g^\prime = \hat g \hat u$ in the standard form (\ref{defh}).
Note that $\hat h$ is a nonlinear function of both $\hat u$ and $\hat g$.
It is sometimes called the compensator.

The covariant derivatives are defined by using the fact that
any variation $\hat g\delta \hat g^\dagger$
is an element of the Lie algebra and can be written as a linear combination
of the generators. The same is true
for $\hat g \left(\partial_\mu - i \hat V_\mu\right)\hat g^\dagger$
if we include external fields
$\hat V_\mu$ transforming as  $\hat V_\mu
\to \hat g \hat V_\mu \hat g^\dagger+i \hat g \partial_\mu \hat g^\dagger$.
We define \cite{CCWZ}
\be
\hat u^\dagger  \left(\partial_\mu - i \hat V_\mu\right) \hat u
\equiv \hat \Gamma_\mu - \frac{i}{2}\hat u_\mu\,,
\qquad
\hat \Gamma_\mu = \Gamma^a_\mu Q^a\,,
\qquad
\hat u_\mu = u_\mu^a X^a\,.
\ee
I.e. $\hat \Gamma_\mu$ is in the conserved part and $\hat u_\mu$ in the
broken part of the Lie algebra.
The transformation under the group $G$ can be derived
from (\ref{defh}) and is
\be
\hat \Gamma_\mu \to \hat h \hat \Gamma_\mu \hat h^\dagger
  +\hat h\partial_\mu \hat h^\dagger\,,\qquad
\hat u_\mu \to \hat h \hat u_\mu \hat h^\dagger\,.
\ee
$\hat u_\mu$ can be used to construct Lagrangians and covariant
derivatives on objects $\psi$ transforming as $\psi\to \hat h\psi$
are defined as
\be
\hat\nabla_\mu \psi = \partial_\mu \psi +\hat \Gamma_\mu \psi\,.
\ee
It can be checked that $\hat\nabla_\mu \psi\to \hat h\hat\nabla_\mu \psi$.
The external fields appear as (axial) vector fields $\hat V_\mu$
and (pseudo) scalar fields $\hat\mathcal{M}$.
The external fields $\hat V_\mu$ show up in $\hat u_\mu$,
covariant derivatives $\hat\nabla_\mu$ and field strengths
$\hat V_{\mu\nu} \equiv \partial_\mu\hat V_\nu-\partial_\nu\hat V_\mu
-i\left[\hat V_\mu,\hat V_\nu\right]$.
The latter can be made to transform simpler by defining the objects
\be
\hat f_{\mu\nu} \equiv \hat u^\dagger \hat V_{\mu\nu} \hat u
\to \hat h \hat f_{\mu\nu} \hat h^\dagger\,.
\ee
$\hat\mathcal{M}\to \hat g \hat\mathcal{M}\hat g^\dagger$
can similarly be made into
\be
\hat\chi\equiv\hat u^\dagger\hat\mathcal{M}\hat u\to
\hat h \hat \chi \hat h^\dagger\,.
\ee
If there exists extra discrete symmetries like parity ($P$)
that leave the unbroken part of the group invariant
objects $O$ like $\hat f_{\mu\nu}$ can be split into
pieces that are independent via $O_\pm\equiv O\pm P(O)$.

In the effective field theory for QCD in terms of $N_F\times N_F$
matrices the notation usually used has the objects
with the associated symmetry transformations:
\ba
\label{QCDstandard}
u&=&\exp\left(\frac{i}{\sqrt{2}\,F}\pi^a T^a\right)
 \to g_R u h^\dagger = h u g_L^\dagger\,,
\nonumber\\
\Gamma_\mu &=&\frac{1}{2}\left(u^\dagger(\partial_\mu-ir_\mu)u
+u(\partial_\mu-l_\mu)u^\dagger\right)
\to h \Gamma_\mu h^\dagger+ih\partial_\mu h^\dagger\,,
\nonumber\\
u_\mu &=&i\left(u^\dagger(\partial_\mu-ir_\mu)u
-u(\partial_\mu-l_\mu)u^\dagger\right)\to h u_\mu h^\dagger\,,
\nonumber\\
\nabla_\mu O &=&\partial_\mu O+\Gamma_\mu O-O\Gamma_\mu
\to h \nabla_\mu O h^\dagger\quad\mathrm{for}\quad O\to h O h^\dagger\,,
\nonumber\\
\chi_\pm &=& u^\dagger \chi u^\dagger\pm u\chi^\dagger u
\to h \chi_\pm h^\dagger\,,
\nonumber\\
f_{\pm\mu\nu}&=& u l_{\mu\nu} u^\dagger \pm u^\dagger r_{\mu\nu} u\,.
\to h f_{\pm\mu\nu} h^\dagger\ea
$l_{\mu\nu}$ and $r_{\mu\nu}$ are the field strengths from $l_\mu$ and $r_\mu$.
$T^a$ are the $SU(N_F)$ generators.
These can be related to the general objects defined in the CCWZ way via
\be
\hat u = \left(\begin{array}{cc} u & 0\\ 0 & u^\dagger\end{array}\right)
\,,\qquad
\hat u_\mu = \left(\begin{array}{cc} u_\mu & 0\\ 0 & -u_\mu\end{array}\right)
\,,\qquad
\hat \Gamma_\mu =
  \left(\begin{array}{cc} \Gamma_\mu & 0\\ 0 & \Gamma_\mu\end{array}\right)
\cdots\,.
\ee
$\chi_\pm$ and $\hat f_{\pm\mu\nu}$ are constructed from $\hat\chi$
and $\hat f_{\mu\nu}$ using parity.
These objects have been used to construct the NLO Lagrangian
and the NNLO Lagrangian \cite{BCE1}.
One of the nontrivial relations used there was
\be
\nabla_\mu u_\nu-\nabla_\nu u_\mu = -f_{-\mu\nu}\,.
\ee
In this notation the lowest order Lagrangian is
\be
\label{QCDLO}
\mathcal{L}_2 = \frac{F^2}{4}\langle u_\mu u^\mu +\chi_+\rangle\,.
\ee
The NLO Lagrangian derived by \cite{GL1} reads (here in the version for
arbitrary $N_F$)
\ba
\label{QCDNLO}
\mathcal{L}_4 &=&
L_0 \langle u^\mu u^\nu u_\mu u_\nu \rangle
+L_1 \langle u^\mu u_\mu\rangle\langle u^\nu u_\nu \rangle
+L_2 \langle u^\mu u^\nu\rangle\langle u_\mu u_\nu \rangle
+L_3 \langle u^\mu u_\mu u^\nu u_\nu \rangle
\nonumber\\&&
+L_4  \langle u^\mu u_\mu\rangle\langle\chi_+\rangle
+L_5  \langle u^\mu u_\mu\chi_+\rangle
+L_6 \langle\chi_+\rangle^2
+L_7 \langle\chi_-\rangle^2
+\frac{1}{2} L_8 \langle\chi_+^2+\chi_-^2\rangle
\nonumber\\&&
-i L_9\langle f_{+\mu\nu}u^\mu u^\nu\rangle
+\frac{1}{4}L_{10}\langle f_+^2-f_-^2\rangle
+H_1\langle l_{\mu\nu}l^{\mu\nu}+r_{\mu\nu}r^{\mu\nu}\rangle
+H_2\langle\chi\chi^\dagger\rangle\,.
\ea

\subsection{Adjoint}
\label{EFTAdjoint}

The vacuum in this case can be characterized by the condensate
\be
\langle \hat q_{i}^T C \hat q_j\rangle
= \frac{1}{2}\langle\overline q_L q_R\rangle
J_{Sij}\,.
\ee
Under the symmetry group $g\in SU(2N_F)$ this moves around as
\be
J_S \to g J_S g^T\,.
\ee
The unbroken part of the group is given by the generators $Q^a$
and the broken part by the generators $X^a$ which satisfy
\be
\label{commutatorsAdjoint}
J_S Q^a = - Q^{aT} J_S\,,\qquad J_S X^a = X^{aT} J_S\,.
\ee
Just as in the QCD case we can now construct a rotated vacuum
in general by using the broken part of the symmetry group
on the vacuum. This leads to a matrix\footnote{In Sect.~\ref{EFTQCD}
we added a hat to many quantities to distinguish the
$N_F\times N_F$ and $2N_F\times 2N_F$ matrices.
This is not needed here and we only keep the hat explicitly
on $\mathcal{M}$.}
\be
U = u J_S u^T\to g U g^T\qquad\mathrm{with}\qquad
u=\exp\left(\frac{i}{\sqrt{2}\,F}\pi^a X^a\right)\,.
\ee
The matrix $u$ transforms as in the general $CCWZ$ case
as
\be
u \to g u h^\dagger\,.
\ee
The earlier work used the matrix $U$ to describe
the Lagrangian \cite{Kogut}. Here we will use the CCWZ scheme
to obtain a notation that is formally identical to the
QCD case. We add full $2N_F\times 2N_F$ matrices of external fields $V_\mu$
and $\hat M$. We need to obtain the $\Gamma_\mu$ and $u_\mu$
parts of $u^\dagger\left(\partial_\mu-iV_\mu\right)u$.
Here several observations are useful.
Eqs.~(\ref{commutatorsAdjoint}) have as a consequence that
matrices like $u$ satisfy
\be
\label{JSu}
u J_S = J_S u^T \,, J_S u = u^T J_S\,.
\ee
A general matrix $F$ can be split two parts,
one behaving as the broken part, the other as the unbroken part
of the group generators. I.e.
\ba
F &=& \overline F +\tilde F\,,
\nonumber\\
\overline F J_S&=& -J_S \overline F^T\,,\qquad \tilde F J_S = \tilde F^T J_S\,,
\nonumber\\
\overline F &=&\frac{1}{2}\left(F-J_S F^T J_S\right)\,,
\nonumber\\
\tilde F &=&\frac{1}{2}\left(F+J_S F^T J_S\right)\,.
\ea
This means that we obtain
\ba
\label{defumuadjoint}
u_\mu &=& i\left(u^\dagger(\partial_\mu-i V_\mu)u
 - u(\partial_\mu+iJ_S V^T_\mu J_S)u^\dagger\right)\,,
\nonumber\\
\Gamma_\mu &=& \frac{1}{2}\left(u^\dagger(\partial_\mu-i V_\mu)u
 + u(\partial_\mu+iJ_SV^T_\mu J_S)u^\dagger\right)\,.
\ea
Here we used the properties (\ref{JSu}).
With these quantities we can construct covariant derivatives and Lagrangians.
The formal similarity to the QCD case is obviously there if we also
use for the vector external fields
\be
l_\mu = -J_S V^T_\mu J_S\,,\qquad r_\mu = V_\mu\,.
\ee
The analogy goes even further since
$v_\mu = r_\mu+l_\mu$ corresponds to the currents from conserved generators
and $a_\mu = r_\mu-l_\mu$ to the currents from the spontaneously broken
generators.
The equivalent quantities to
the field strengths are
\be
\label{deffieldadjoint}
f_{\pm\mu\nu} = J_S u V_{\mu\nu} u^\dagger J_S\pm  u V_{\mu\nu} u^\dagger
\ee
with $V_{\mu\nu}=\partial_\mu V_\nu-\partial_\nu V_\mu-i
\left(V_\mu V_\nu- V_\nu V_\mu\right)$ and for the mass matrix
\ba
\label{defchiadjoint}
\chi_\pm &=& u^\dagger\chi u^{\dagger T} J_S \pm J_S u^T\chi^\dagger u
\nonumber\\
&=&
 u^\dagger\chi J_S u^{\dagger}  \pm  u J_S\chi^\dagger u\,,
\ea
with $\chi = 2B_0 \hat\mathcal{M}$.
The Lagrangians at LO and NLO have exactly the same
form as (\ref{QCDLO}) and (\ref{QCDNLO}) but now with
$u_\mu$, $\chi_\pm$ and $f_{\pm\mu\nu}$
as defined in (\ref{defumuadjoint}), (\ref{deffieldadjoint})
and (\ref{defchiadjoint}).

\subsection{Two colours}
\label{EFT2colour}

The vacuum in this case can be characterized by the condensate
$
\langle \hat q_{\alpha i}^T C \epsilon_{\alpha\beta}\hat q_{\beta j}\rangle
= \frac{1}{2}\langle\overline q_L q_R\rangle
J_{Aij}\,.
$ 
Under the symmetry group $g\in SU(2N_F)$ this moves around as
$
J_A \to g J_A g^T\,.
$
The unbroken part of the group is given by the generators $Q^a$
and the broken part by the generators $X^a$ which satisfy
$
J_A Q^a = - Q^{aT} J_A\,,~ J_A X^a = X^{aT} J_A\,.
$ 
Just as in the QCD and the adjoint case we construct a rotated vacuum
by using the broken part of the symmetry group
on the vacuum. This leads to a matrix\footnote{
The formulas in this subsection are almost identical
with those in the previous subsection but $J_A^2=-1$
while $J_S^2=1$. We have put in those by introducing $J_A^T$
rather than $J_A$ in a few places.}
$
U = u J_A u^T\to g U g^T$ with
$u=\exp\left(\frac{i}{\sqrt{2}\,F}\pi^a X^a\right)\,.
$
The matrix $u$ transforms as
$
u \to g u h^\dagger\,.
$
Ref.~\cite{Kogut} used the matrix $U$ to describe
the Lagrangian. Here we use the CCWZ scheme.
We add full $2N_F\times 2N_F$ matrices of external fields $V_\mu$
and $\hat\mathcal{M}$ and then need to obtain the $\Gamma_\mu$ and $u_\mu$
parts of $u^\dagger\left(\partial_\mu-iV_\mu\right)u$.
Matrices like $u$ satisfy $u J_A = J_A u^T$ and $J_A u = u^T J_A$.

A general matrix $F$ can be split two parts,
one behaving as the broken part, the other as the unbroken part
of the group generators. I.e.
\ba
F &=& \overline F +\tilde F\,,
\qquad\qquad
\overline F J_A = -J_A \overline F^T\,,\qquad \tilde F J_A^T
 = \tilde F^T J_A\,,
\nonumber\\
\overline F &=&\frac{1}{2}\left(F-J_A F^T J_A^T\right)\,,
\qquad
\tilde F =\frac{1}{2}\left(F+J_A F^T J_A^T\right)\,.
\ea
Using this, we obtain
\ba
\label{defumu2colour}
u_\mu &=& i\left(u^\dagger(\partial_\mu-i V_\mu)u
 - u(\partial_\mu+iJ_A V^T_\mu J_A^T)u^\dagger\right)\,,
\nonumber\\
\Gamma_\mu &=& \frac{1}{2}\left(u^\dagger(\partial_\mu-i V_\mu)u
 + u(\partial_\mu+iJ_AV^T_\mu J_A^T)u^\dagger\right)\,.
\ea
Covariant derivatives and Lagrangians are constructed as above.
The formal similarity to the QCD case is once more obviously if we
use for the vector external fields
\be
l_\mu = -J_A V^T_\mu J_A^T\,,\qquad r_\mu = V_\mu\,.
\ee
Again
$v_\mu = r_\mu+l_\mu$ corresponds to the currents from conserved generators
and $a_\mu = r_\mu-l_\mu$ to the currents from the spontaneously broken
generators.
The equivalent quantities to
the field strengths are
\be
\label{deffield2colour}
f_{\pm\mu\nu} = J_A u V_{\mu\nu} u^\dagger J_A^T\pm  u V_{\mu\nu} u^\dagger
\ee
with $V_{\mu\nu}=\partial_\mu V_\nu-\partial_\nu V_\mu-i
\left(V_\mu V_\nu- V_\nu V_\mu\right)$ and for the mass matrix
\ba
\label{defchi2colour}
\chi_\pm &=& u^\dagger\chi u^{\dagger T} J_A^T \pm J_A u^T\chi^\dagger u
= u^\dagger\chi J_A^T u^{\dagger}  \pm  u J_A\chi^\dagger u\,,
\ea
with $\chi = 2B_0 \hat\mathcal{M}$.
The Lagrangians at LO and NLO have exactly the same
form as (\ref{QCDLO}) and (\ref{QCDNLO}) but with
$u_\mu$, $\chi_\pm$ and $f_{\pm\mu\nu}$ as defined in this subsection.

\section{The divergence structure at NLO}
\label{infinities}

When going beyond tree level renormalization becomes
necessary. A thorough discussion of renormalization in ChPT
at NNLO can be found in \cite{BCE2,BCEGS}. We use here the same conventions
and subtraction procedure.
This means that the NLO LECs are replaced by
\be
\label{defLir}
L_i = \left(c \mu\right)^{d-4}
\left[\Gamma_i\Lambda+L_i^r(\mu)\right]\,,
\ee
with $\Lambda = 1/(16\pi^2(d-4))$ and
$\ln c = -[\ln 4\pi+\Gamma^\prime(1)+1]/2$.
The constants $\Gamma_i$ were calculated for the $QCD$ case
in \cite{GL2}. The same method can be generalized to the case here.
The calculation is extremely similar for all three cases.
The method is the same as the one in \cite{GL2}.
We split $u$ in a classical and a quantum part
\be
u = u_c e^{i\xi}\quad\mathrm{with}\quad \xi = \sum_a \xi^a X^a\,.
\ee
The second variation w.r.t. $\xi$ of the LO Lagrangian
can be rewritten in the form
\be
\mathcal{L} = \frac{F^2}{2}\left(d_\mu\xi^a d^\mu\xi^a-
\xi^a \tilde\sigma^{ab}\xi^b\right)\,,
\ee
with $d_\mu\xi^a = \partial_\mu\xi^a+\tilde\Gamma_\mu^{ab}\xi^b$.
The divergence at one-loop level is given by \cite{GL2}
\be
\label{divergence}
-\frac{1}{16\pi^2(d-4)}
\left(\frac{1}{12}\tilde\Gamma_{\mu\nu}^{ab}\tilde\Gamma^{ba\mu\nu}
+\frac{1}{2}\tilde\sigma^{ab}\tilde\sigma{ba}\right)\,.
\ee
Notice that the indices here run over the broken generators
and $\tilde\Gamma_{\mu\nu}^{ab}
=\partial_\mu\tilde\Gamma^{ab}_\nu-\partial_\nu\tilde\Gamma^{ab}_\mu
+\tilde\Gamma_\mu^{ac}\tilde\Gamma_\nu^{cb}
-\tilde\Gamma_\nu^{ac}\tilde\Gamma_\mu^{cb}$.

The expansion for all three cases is identical and leads to
\ba
\tilde\Gamma_\mu^{ab} &=& -\trf{[X^a,X^b]\Gamma_\mu}\,,
\nonumber\\
\tilde\sigma^{ab} &=&
 -\frac{1}{8}\trf{\{X^a,X^b\}\left(\chi_+ +u_\mu u^\mu\right)}
 +\frac{1}{2}\trf{X^a u_\mu X^b u^\mu}\,.
\ea
The difficulty in evaluating (\ref{divergence}) is now rewriting the
sums over broken generators into traces over the original
matrices $u_\mu,\ldots$. In the QCD case, the $X^a$ are
$SU(N_F)$ generators and one can use the formulas
with the $\trf{A}$ going from $1,\ldots,N_F$.
\ba
\mathrm{QCD:}\qquad&&\nonumber\\
\trf{X^a A X^a B}&=& \trf{A}\trf{B}
 -\frac{1}{N_F}\trf{AB}\,,
\nonumber\\
\trf{ X^a A}\trf{X^a B}&=&
 \trf{AB}
 -\frac{1}{N_F}\trf{A}\trf{B}\,.
\ea
There exist similar formulas for the adjoint case
now with  $\trf{A}$ going from $1,\ldots,2N_F$.
\ba
\mathrm{Adjoint:}\qquad&&\nonumber\\
\trf{X^a A X^a B}&=& \frac{1}{2}\trf{A}\trf{B}
 +\frac{1}{2}\trf{AJ_S B^T J_S}-\frac{1}{2N_F}\trf{AB}\,,
\nonumber\\
\trf{ X^a A}\trf{X^a B}&=&
 \frac{1}{2}\trf{AB}+\frac{1}{2}\trf{AJ_S B^T J_S}
 -\frac{1}{2N_F}\trf{A}\trf{B}\,.
\ea
The equivalent formula for the two-colour case is \cite{Splittorff},
again with
 $\trf{A}$ going from $1,\ldots,2N_F$.
\ba
\mathrm{2-colour:}\qquad&&\nonumber\\
\trf{X^a A X^a B}&=& \frac{1}{2}\trf{A}\trf{B}
+\frac{1}{2}\trf{AJ_A B^T J_A} -\frac{1}{2N_F}\trf{AB}\,,
\nonumber\\
\trf{ X^a A}\trf{X^a B}&=&
 \frac{1}{2}\trf{AB}-\frac{1}{2}\trf{AJ_A B^T J_A}
 -\frac{1}{2N_F}\trf{A}\trf{B}\,.
\ea
In all three cases these lead to
\be
\tilde\Gamma_{\mu\nu}^{ab}
= -\trf{[X^a,X^b]\Gamma_{\mu\nu}}\,.
\ee

Repetitive use of these identities allows to rewrite (\ref{divergence})
in the form of (\ref{QCDNLO}). These divergences are then absorbed into
the redefinition of the NLO LECs (\ref{defLir}). The needed
constants $\Gamma_i$ for the three cases are given in
Tab.~\ref{tabdivergence}. We agree with \cite{GL2} for the QCD case,
have a small discrepancy with \cite{Splittorff} for the
two-colour case, our coefficients for $\Gamma_0$ are $\Gamma_3$ are different.
The adjoint case is obtained here for the first time.
\begin{table}
\begin{center}
\begin{tabular}{|c|c|c|c|}
\hline
i & QCD & Adjoint & 2-colour\\
\hline
0  & $N_F/48$              & $(N_F+4)/48$            & $(N_F-4)/48$           \\
1  & $1/16$                & $1/32$                  & $1/32$                 \\
2  & $1/8$                 & $1/16$                  & $1/16$                 \\
3  & $N_F/24$              & $(N_F-2)/24$            & $(N_F+2)/24$           \\
4  & $1/8$                 & $1/16$                  & $1/16$                 \\
5  & $N_F/8$               & $N_F/8$                 & $N_F/8$                \\
6  & $(N_F^2+2)/(16 N_F^2)$& $(N_F^2+1)/(32 N_F^2)$  & $(N_F^2+1)/(32 N_F^2)$ \\
7  & 0                     & 0                       & 0                      \\
8  & $(N_F^2-4)/(16N_F)$   & $(N_F^2+N_F-2)/(16N_F)$ & $(N_F^2-N_F-2)/(16N_F)$\\
9  & $N_F/12$              & $(N_F+1)/2$             & $(N_F-1)/2$            \\
10 & $-N_F/12$             & $-(N_F+1)/2$            & $-(N_F-1)/2$           \\
1' & $-N_F/24$             & $-(N_F+1)/4$            & $-(N_F+1)/4$           \\
2' & $(N_F^2-4)/(8N_F)$    & $(N_F^2+N_F-2)/(8N_F)$  & $(N_F^2-N_F-2)/(8N_F)$ \\
\hline
\end{tabular}
\end{center}
\caption{\label{tabdivergence} The coefficients $\Gamma_i$
for the three cases that are needed to absorb the divergences at NLO.
The last two lines correspond to the terms with $H_1$ and $H_2$.}
\end{table}

\section{The calculation: mass, decay constant and condensate}
\label{NNLO}

In this section we calculate the corrections to the
vacuum expectation value, the meson mass and the decay constant.
The calculations in the work on three-flavour ChPT were done
using FORM \cite{FORM}
and in the loops an explicit sum over all possible particles
was always implemented. For this work we have rewritten the flavour routines
used in that work to use a general sum over the flavour indices
and since we always calculate in the case where
$\mathcal{M} = \mathrm{diag}(\hat m,\ldots,\hat m)$ we then use
the trace formulas of the previous section to perform the sum.

We have checked that our calculations reproduce all the known results
and for the QCD case that all infinities cancel when the NNLO divergence of
\cite{BCE2} is used. For the adjoint and two-colour case we observe that
the nonlocal divergence cancels as it should.

\begin{figure}
\begin{center}
\includegraphics[width=0.8\textwidth]{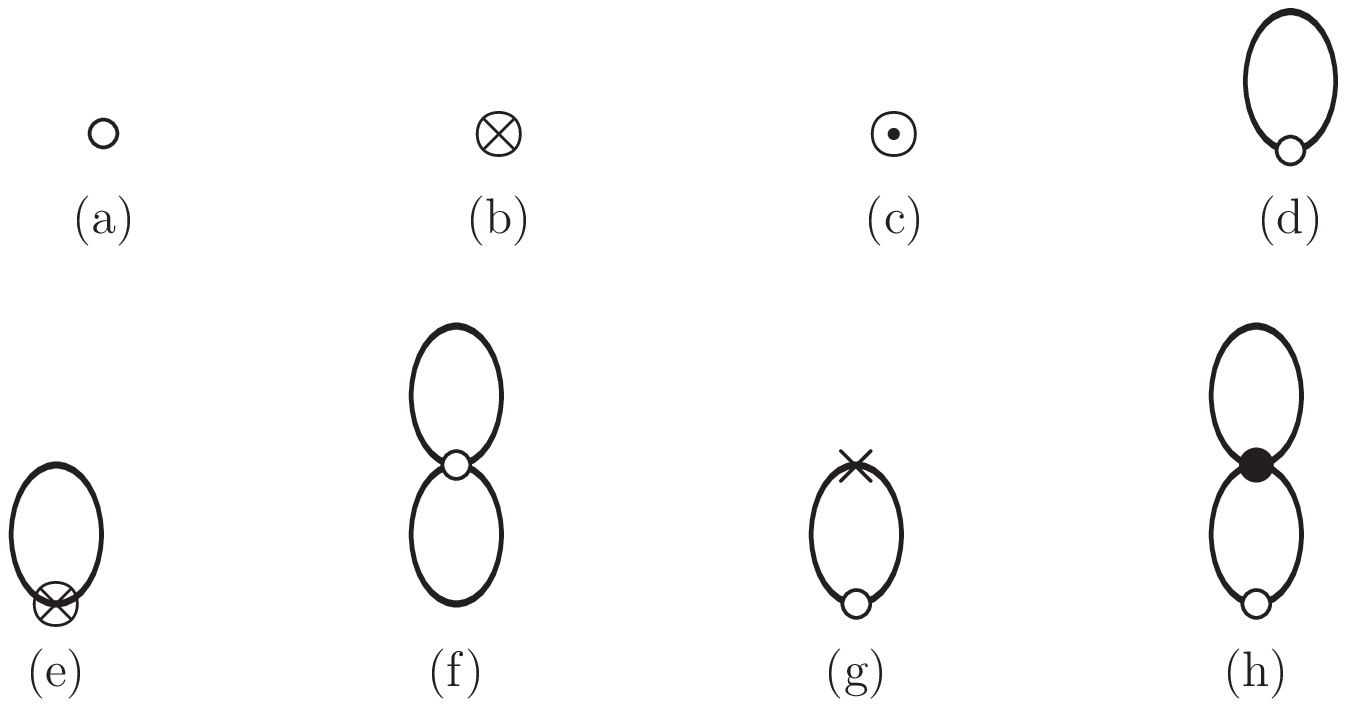}
\end{center}
\caption{\label{figqbarqdiagrams} The diagrams up to order $p^6$ for
$\langle\overline qq\rangle$. The lines are meson propagators and the vertices
are:
$\circ$ a $p^2$ insertion of $\overline qq$,
$\otimes$ a $p^4$ insertion of $\overline qq$,
$\odot$ a $p^6$ insertion of $\overline qq$,
$\bullet$ a $p^2$ vertex and
$\times$ a $p^4$ vertex.}
\end{figure}

The diagrams for the vacuum expectation value are shown in
Fig.~\ref{figqbarqdiagrams}.
The lowest order is the same for all three cases
\be
\langle \overline q q\rangle_\mathrm{LO} \equiv \sum_{i=1,N_F}
\langle \overline q_{Ri}q_{Li}+\overline q_{Li}q_{Ri}\rangle_\mathrm{LO}
= -N_F B_0 F^2\,.
\ee
We use $M^2$ as notation for the lowest order meson mass
\be
\label{defLOmass}
M^2 = 2 B_0\hat m
\ee
and in addition the function
\be
\overline A(M^2) = -\frac{M^2}{16\pi^2}\log\frac{M^2}{\mu^2}\,.
\ee
The integrals needed at the two-loop level are evaluated with the
methods of \cite{GS}
and they can all be expressed in terms of $\overline A(M^2)$.

We express the final result
as
\be
\langle \overline q q\rangle =
\langle \overline q q\rangle_\mathrm{LO}+
\langle \overline q q\rangle_\mathrm{NLO}+
\langle \overline q q\rangle_\mathrm{NNLO}\,.
\ee
The individual parts can be written in terms of logarithms and
analytic contributions as
\ba
\label{expansionqbarq}
\langle \overline q q\rangle_\mathrm{NLO}
 &=& \langle \overline q q\rangle_\mathrm{LO}
\left(a_V \frac{\overline A(M^2)}{F^2}+b_V\frac{M^2}{F^2}\right)\,,
\nonumber\\
\langle \overline q q\rangle_\mathrm{NNLO}
 &=& \langle \overline q q\rangle_\mathrm{LO}
\Bigg(c_V \frac{\overline A(M^2)^2}{F^4}
     +\frac{M^2\overline A(M^2)}{F^4}\left(d_V+\frac{e_V}{16\pi^2}\right)
\nonumber\\
  &&\qquad   +\frac{M^4}{F^4}\left(f_V+\frac{g_V}{16\pi^2}\right)\Bigg)\,.
\ea
The coefficients for the three cases are given in Tab.~\ref{tabqbarq}.
Note that we use the same notation for the LECs in the
three cases but they are different LECs and in addition different for
different values of $N_F$.
\begin{table}
\begin{center}
\begin{tabular}{|c|c|}
\hline
 & QCD \\
\hline
$a_V$ & $N_F-\frac{1}{N_F}$ \\
$b_V$ & $16 N_F L_6^r+8 L_8^r +4 H_2^r$\\
$c_V$ & $\frac{3}{2}\left(-1+\frac{1}{N_F^2}\right)$\\
$d_V$ & $-24\left(N_F^2-1\right)\,
  \left(L_4^r-2L_6^r+\frac{1}{N_F}(L_5^r-2L_8^r)\right)$\\
$e_V$ & $1-\frac{1}{N_F^2}$\\
$f_V$ & $48\left(K_{25}^r+N_F K_{26}^r+N_F^2 K_{27}^r\right)$\\
$g_V$ & $8\left(N_F^2-1\right)\,
\left(L_4^r-2L_6^r+\frac{1}{N_F}(L_5^r-2L_8^r)\right)$\\
\hline
 & Adjoint \\
\hline
$a_V$ & $ N_F+\frac{1}{2}-\frac{1}{2N_F}$\\
$b_V$ & $32 N_F L_6^r+8 L_8^r +4 H_2^r$\\
$c_V$ & $\frac{3}{8}\left(-1+\frac{1}{N_F^2}-\frac{2}{N_F}+2N_F\right)$\\
$d_V$ & $-12\left(2N_F^2+N_F-1\right)
  \left(2L_4^r-4L_6^r+\frac{1}{N_F}(L_5^r-2L_8^r)\right)$\\
$e_V$ & $\frac{1}{4}\left(1-\frac{1}{N_F^2}+\frac{2}{N_F}-2N_F\right)$\\
$f_V$ & $r^r_{VA}$ \\
$g_V$ & $4\left(2N_F^2+N_F-1\right)
  \left(2L_4^r-4L_6^r+\frac{1}{N_F}(L_5^r-2L_8^r)\right)$\\
\hline
 & 2-colour\\
\hline
$a_V$ & $ N_F-\frac{1}{2}-\frac{1}{2N_F}$\\
$b_V$ & $32 N_F L_6^r+8 L_8^r +4 H_2^r$ \\
$c_V$ & $\frac{3}{8}\left(-1+\frac{1}{N_F^2}+\frac{2}{N_F}-2N_F\right)$ \\
$d_V$ &  $-12\left(2N_F^2-N_F-1\right)
  \left(2L_4^r-4L_6^r+\frac{1}{N_F}(L_5^r-2L_8^r)\right)$\\
$e_V$ & $\frac{1}{4}\left(1-\frac{1}{N_F^2}-\frac{2}{N_F}+2N_F\right)$ \\
$f_V$ & $r^r_{VT}$ \\
$g_V$ & $4\left(2N_F^2-N_F-1\right)
  \left(2L_4^r-4L_6^r+\frac{1}{N_F}(L_5^r-2L_8^r)\right)$ \\
\hline
\end{tabular}
\end{center}
\caption{\label{tabqbarq} The coefficients $a_V,\ldots,g_V$
appearing in the expansion of the vacuum expectation value.}
\end{table}
The infinite parts can be absorbed in the NNLO Lagrangian
coefficients by writing
\ba
r_i = \left(c\mu\right)^{2(d-4)}
\left(r^r_i-\Gamma_i^{(2)}\Lambda^2
-\left(\frac{1}{16\pi^2}\Gamma_i^{(1)}+\Gamma_i^{(L)}
\right)\Lambda
\right)\,.
\ea
The subtractions needed for the QCD case have been derived in general
before in \cite{BCE2}. The adjoint and two-colour case can be made finite
by the following:
\ba
\Gamma^{(2)}_{VA} &=& \frac{3}{2}
    \left(1-\frac{1}{N_F^2}+2\frac{1}{N_F}-2 N_F\right)\,,
\nonumber\\
\Gamma^{(L)}_{VA} &=&24\left(2N_F^2+N_F-1\right)\,
  \left(2L_4^r-4L_6^r+\frac{1}{N_F}(L_5^r-2L_8^r)\right)\,,
\nonumber\\
\Gamma^{(1)}_{VA} &=& 0\,,
\nonumber\\
\Gamma^{(2)}_{VT} &=& \frac{3}{2}
    \left(1-\frac{1}{N_F^2}-2\frac{1}{N_F}+2 N_F\right)\,,
\nonumber\\
\Gamma^{(L)}_{VT} &=&24\left(2N_F^2-N_F-1\right)\,
  \left(2L_4^r-4L_6^r+\frac{1}{N_F}(L_5^r-2L_8^r)\right)\,,
\nonumber\\
\Gamma^{(1)}_{VT} &=& 0\,.
\ea
This result agrees at NLO with \cite{GL4} for the QCD case
and \cite{Splittorff}\footnote{Those authors used a different normalization
for $F$. Ours corresponds to $F_\pi\approx 93$~MeV for the QCD case and $N_c=3$.}
for the 2-colour case. It also agrees for $N_F=3$
at NNLO with \cite{ABT3,BG}. The remaining results are new.

\begin{figure}
\begin{center}
\includegraphics[width=0.8\textwidth]{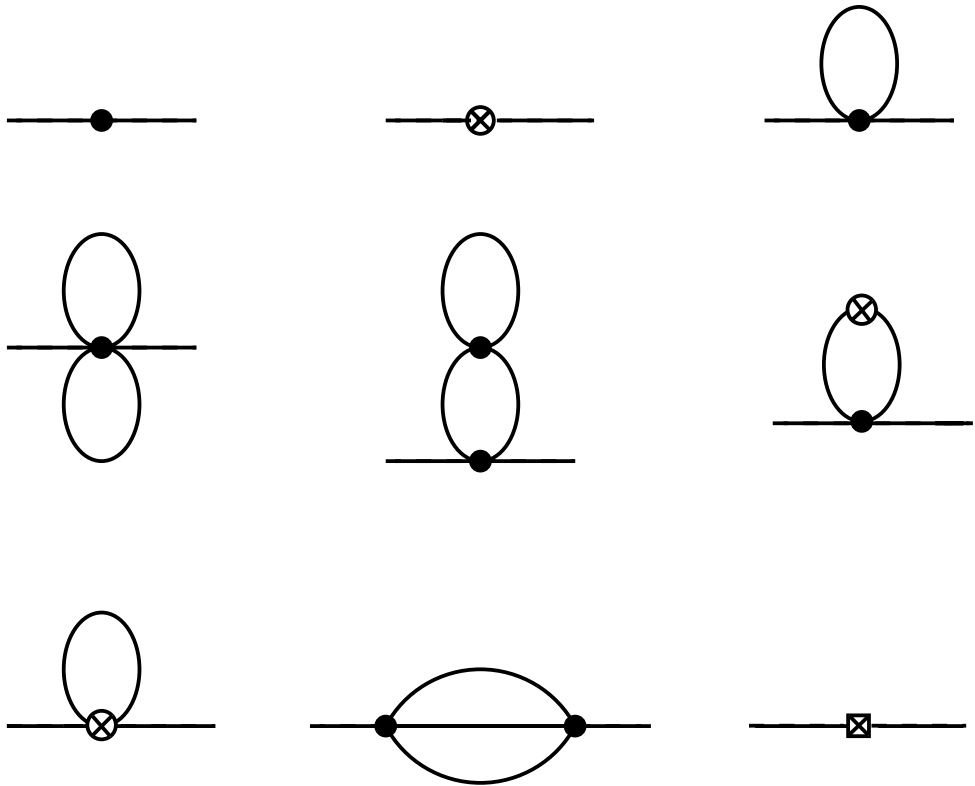}
\end{center}
\caption{\label{figmassdiagrams} The diagrams up to order $p^6$ for
the meson self energy. The lines are meson propagators and the vertices
are:
$\bullet$ a $p^2$ vertex,
$\times$ a $p^4$ vertex and a crossed box a $p^6$ vertex.
The diagrams for the decay constant are the same with one external meson leg
replaced by an axial current.}
\end{figure}

We perform the expansion of the physical meson mass
to the same order. The physical mass can be written as
\be
M^2_\mathrm{phys} = M^2_\mathrm{LO}
+ M^2_\mathrm{NLO}+M^2_\mathrm{NNLO}\,.
\ee
The lowest order was already given in (\ref{defLOmass}) and is the same for all
three cases.
The two higher order can be expanded in logarithms and analytical contributions
via
\ba
\label{expansionmass}
M^2_\mathrm{NLO}
 &=& M^2
\left(a_M \frac{\overline A(M^2)}{F^2}+b_M\frac{M^2}{F^2}\right)\,,
\nonumber\\
M^2_\mathrm{NNLO}
 &=& M^2
\Bigg(c_M \frac{\overline A(M^2)^2}{F^4}
     +\frac{M^2\overline A(M^2)}{F^4}\left(d_M+\frac{e_M}{16\pi^2}\right)
\nonumber\\
 &&\qquad
     +\frac{M^4}{F^4}\left(f_M+\frac{g_M}{16\pi^2}+\frac{h_M}{(16\pi^2)^2}
     \right)\Bigg)\,.
\ea
The mass can be calculated by finding the zeros of the inverse propagator,
see e.g. the discussion \cite{ABT1}. The relevant one-particle irreducible
diagrams are shown in Fig.~\ref{figmassdiagrams}.
The coefficients for the three cases are given in Tab.~\ref{tabmass}.
\begin{table}
\begin{center}
\begin{tabular}{|c|c|}
\hline
 & QCD \\
\hline
$a_M$ & $-\frac{1}{N_F}$ \\
$b_M$ & $8 N_F\left(2 L_6^r- L_4^r\right)+8 \left(2L_8^r-L_5^r\right)$\\
$c_M$ & $-\frac{1}{2}+\frac{9}{2N_F^2}+\frac{3}{8}N_F^2$\\
$d_M$ & $8L_0^r( -{3\over N_F}+N_F)+8 L_1^r(-1+2N^2_F)+4L_2^r(4+N^2_F)+ L_3^r(-{24\over N_F}+20N_F) $\\
      & $+L_4^r( 40-16N^2_F)+ L_5^r( {40\over N_F}-16N_F)+ L_6^r(-16+16N^2_F)+L_8^r(-{80\over N_F}+32N_F)$\\
$e_M$ & $-{5\over3}+{4\over N^2_F}+{19\over16}N^2_F$\\
$f_M$ & $-32 K_{17}^r-16K_{19}^r-16K_{23}^r+48K_{25}^r+32K_{39}^r$\\
      & $+N_F\left(-32 K_{18}^r-16K_{20}^r-16K_{21}^r
         +48K_{26}^r+32K_{40}^r\right)$\\
      & $  +N^2_F\left(-16K_{22}^r+48K_{27}^r\right)
        +64(N_F L^r_4+L^r_5)(N_F L^r_4+L^r_5 - 2N_F L^r_6-2L^r_8)$\\
$g_M$ & $-{4\over N_F}(L_0^r+L_3^r) + 4L_1^r +2N_F(2L_0^r+L_3^r)+2N^2_FL_2^r$\\
      & $-8[L^r_4-2L^r_6+{1\over N_F}(L^r_5-2L^r_8)]$\\
$h_M$ & $-\frac{1}{4}+\frac{3}{4}\frac{1}{N_F^2}+\frac{169}{384}N_F^2$\\
\hline
 & Adjoint \\
\hline
$a_M$ & ${1\over2}-\frac{1}{2N_F}$\\
$b_M$ & $16N_F(2L_6^r-L^r_4)+8( 2L_8^r-L_5^r) $\\
$c_M$ & $\frac{3}{8}\left(1+\frac{3}{N_F^2}-\frac{4}{N_F}+N_F+N^2_F\right)$\\
$d_M$ & $L_0^r( 12-12{1\over N_F}+8N_F)+8L_1^r(-1+2N_F+4N^2_F)$\\
      & $+4L_2^r(4+N_F+2N^2_F)+ L_3^r(12-{12\over N_F}+20N_F) $\\
      & $+L_4^r( 40-40N_F-32N^2_F)+L_5^r(-20+{20\over N_F}-16N_F)$\\
      & $+ 16L_6^r(-1+3N_F+2N^2_F)+ L_8^r(40-{40\over N_F}+32N_F) $\\
$e_M$ & $-{2\over3}+{1\over N^2_F}-{3\over4}{1\over N_F}+{77\over48}N_F+{19\over16}N^2_F$\\
$f_M$ & $r^r_{MA}+64(2N_F L^r_4+L^r_5)(2N_F L^r_4+L^r_5-4N_F L^r_6-2L^r_8)$ \\
$g_M$ & $2L_0^r( 1-{1\over N_F}+2N_F)+4 L_1^r+2N_FL_2^r(1+2N_F)+ 2L_3^r(1-{1\over N_F}+N_F)$\\
      & $-8( 1-N_F)(L_4^r-2L_6^r)+4(1-{1\over N_F})(L_5^r-2L_8^r) $\\
$h_M$ & $-\frac{1}{16}+\frac{3}{16}\frac{1}{N_F^2}-\frac{3}{16}\frac{1}{N_F}
     +\frac{193}{384}N_F+\frac{169}{384}N_F^2$\\
\hline
 & 2-colour\\
\hline
$a_M$ & $-{1\over2}-\frac{1}{2N_F}$\\
$b_M$ & $16N_F(2L_6^r-L^r_4)+8( 2L_8^r-L_5^r) $\\
$c_M$ & $\frac{3}{8}\left(1+\frac{3}{N_F^2}+\frac{4}{N_F}-N_F+N^2_F\right)$\\
$d_M$ & $L_0^r( -12-12{1\over N_F}+8N_F)+8 L_1^r(-1-2N_F+4N^2_F)$\\
      & $+4L_2^r(4-N_F+2N^2_F)+ L_3^r(-12-{12\over N_F}+20N_F) $\\
      & $+L_4^r( 40+40N_F-32N^2_F)+L_5^r(20+{20\over N_F}-16N_F)$\\
      & $+ 16L_6^r(-1-3N_F+2N^2_F)+ L_8^r(-40-{40\over N_F}+32N_F) $\\
$e_M$ & $-{2\over3}+{1\over N^2_F}+{3\over4}{1\over N_F}-{77\over48}N_F+{19\over16}N^2_F$\\
$f_M$ & $r^r_{MT}+64(2N_F L^r_4+L^r_5)(2N_F L^r_4+L^r_5-4N_F L^r_6-2L^r_8)$ \\
$g_M$ & $-2L_0^r( 1+{1\over N_F}-2N_F)+4 L_1^r-2N_FL_2^r(1-2N_F) - 2L_3^r(1+{1\over N_F}-N_F)$\\
      & $-8( 1+N_F)(L_4^r-2L_6^r)-4(1+{1\over N_F})(L_5^r-2L_8^r) $\\
$h_M$ & $-\frac{1}{16}+\frac{3}{16}\frac{1}{N_F^2}+\frac{3}{16}\frac{1}{N_F}
-\frac{193}{384}N_F+\frac{169}{384}N_F^2$\\
\hline
\end{tabular}
\end{center}
\caption{\label{tabmass} The coefficients $a_M,\ldots,g_M$
appearing in the expansion of the mass.}
\end{table}
The subtractions needed for the QCD case have been derived in
general before in \cite{BCE2}. The adjoint and two-colour case can
be made finite by the following:
\ba
\Gamma_{MA}^{(2)}&=&{1\over2}\left(1- {9\over N_F^2} + {12\over
N_F}-7N_F-3N_F^2\right)\,,
\nonumber\\
\Gamma_{MA}^{(L)}&=& -8 \Bigg[ \left(3 -{3\over N_F}+ 2N\right)L^r_0
+2\left(-1 + 2N_F+ 4N_F^2\right)L^r_1 +\left(4 + N_F+
2N_F^2\right)L^r_2
\nonumber\\
&&+\left(3 -{3\over N_F}+ 5N_F\right)L^r_3 +{2\over N_F}(2 - 2N_F-
3N_F^2)(2N_FL^r_4 +L^r_5)
\nonumber\\
&& +4\left(-1 + 3N_F+ 4N_F^2\right)L^r_6+\left(10 - {10\over N_F}+
12N_F\right)L^r_8)\Bigg]\,,
\nonumber\\
\Gamma_{MA}^{(1)}&=&-{1\over4}\left(-{5\over3}+{5\over N_F^2}-
{5\over
N_F}+{67\over 12}N_F+{47\over 12}N_F^2\right)\,,\nonumber\\
 \Gamma^{(2)}_{MT} &=& {1\over2}\left(1- {9\over N_F^2} - {12\over
N_F}+7N_F-3N_F^2\right)\,,
\nonumber\\
\Gamma^{(L)}_{MT} &=&-8 \Bigg[
\left(-3 -{3\over N_F}+ 2N_F\right)L^r_0
+2\left(-1 - 2N_F+ 4N_F^2\right)L^r_1+ \left(4 - N_F+ 2N_F^2\right)L^r_2
\nonumber\\
&&+\left(-3 -{3\over N_F}+ 5N_F\right)L^r_3
 +{2\over N_F}(2 + 2N_F- 3N_F^2)(2N_FL^r_4+L^r_5)
\nonumber\\
&&+4\left(-1 - 3N_F+ 4N_F^2\right)L^r_6+\left(-10 -\frac{10}{N_F}+
12N_F\right)L^r_8)\Bigg]
\nonumber\,,\\
\Gamma^{(1)}_{MT} &=&-{1\over4}\left(-{5\over3}+{5\over N_F^2}
+{5\over N_F}-{67\over 12}N_F+{47\over 12}N_F^2\right)\,.
\ea
This
result agrees at NLO with \cite{GL4} for the QCD case and
\cite{Splittorff} for the 2-colour case. It also agrees with the
masses for two and three flavours in the QCD case as calculated in
\cite{Buergi,BCEGS,ABT1,GK}. The remaining results are new.

We perform the expansion of the physical decay constant
to the same order. The decay constant can be written as
\be
F_\mathrm{phys} = F_\mathrm{LO}
+ F_\mathrm{NLO}+F_\mathrm{NNLO}\,.
\ee
The lowest order is $F_\mathrm{LO}=F$ and is the same for all
three cases.
The two higher order can be expanded in logarithms and analytical contributions
via
\ba
\label{expansiondecay}
F_\mathrm{NLO}
 &=& F
\left(a_F \frac{\overline A(M^2)}{F^2}+b_F\frac{M^2}{F^2}\right)\,,
\nonumber\\
F_\mathrm{NNLO}
 &=& F
\Bigg(c_F \frac{\overline A(M^2)^2}{F^4}
     +\frac{M^2\overline A(M^2)}{F^4}\left(d_F+\frac{e_F}{16\pi^2}\right)
\nonumber\\
&&\qquad     +\frac{M^4}{F^4}\left(f_F+\frac{g_F}{16\pi^2}+\frac{h_F}{(16\pi^2)^2}
      \right)\Bigg)\,.
\ea
The decay constant can be calculated by computing the
one-meson matrix element of the axial current.
The diagrams for the wave-function renormalization are the same
as those for the mass in Fig.~\ref{figmassdiagrams} and those
for the bare matrix-element are again those of Fig.~\ref{figmassdiagrams}
but with one external meson leg replaced by the axial current.
The coefficients for the three cases are given in Tab.~\ref{tabdecay}.
\begin{table}
\begin{center}
\begin{tabular}{|c|c|}
\hline
 & QCD \\
\hline
$a_F$ & $\frac{1}{2}N_F$ \\
$b_F$ & $4 N_F L_4^r+4 L_5^r$\\
$c_F$ & $-\frac{1}{2}-\frac{3}{16}N_F^2$\\
$d_F$ & $\frac{4}{N_F} (3 L_0^r+3 L_3^r- L_5^r)
     +4 L_1^r-8 L_2^r-4L_4^r
    +N_F (-4 L_0^r-10 L_3^r-2 L_5^r+8 L_8^r)$\\
      &$    +2N_F^2 (-4 L_1^r- L_2^r- L_4^r+4L_6^r) $\\
$e_F$ & $\frac{2}{3}-\frac{1}{2N_F^2}-\frac{59}{96}N_F^2$\\
$f_F$ &$-8\left(N_F L_4^r+L_5^r\right)^2+
8 (K_{19}^r+ K_{23}^r)+8 N_F (K_{20}^r+K_{21}^r)+8 N_F^2 K_{22}^r$\\
$g_F$ & $\frac{2}{N_F}(L_0^r+L_3^r)
     -2L_1^r+N_F (-2 L_0^r-L_3^r+4L_5^r-8 L_8^r)
   +N_F^2 (-L_2^r+4L_4^r-8L_6^r) $\\
$h_F$ & $-\frac{7}{24}+\frac{7}{8N_F^2}+\frac{1}{768}N_F^2$\\
\hline
 & Adjoint \\
\hline
$a_F$ & $\frac{1}{2}N_F$ \\
$b_F$ & $8 N_F L_4^r+4 L_5^r$\\
$c_F$ & $-\frac{1}{4}+\frac{3}{16}N_F-\frac{3}{16}N_F^2$\\
$d_F$ &$L_0^r(-6 +{6\over N_F}-4N_F)+4L_1^r(1-2N_F-4N^2_F)-2L_2^r(4+N_F+2N^2_F)$
\\
      & $+ L_3^r(-6+{6\over N_F}-10N_F)
       -4L_4^r( 1-N_F+N^2_F)+2 L_5^r(1- {1\over N_F}-N_F)$\\
      & $+ 8N_F(2N_F L_6^r+L_8^r)$\\
$e_F$ & $\frac{7}{24}-\frac{1}{8N_F^2}+\frac{1}{8N_F}-\frac{29}{32}N_F-\frac{59}{96}N_F^2$\\
$f_F$ & $r^r_{FA}-8(2N_F L_4^r+L_5^r)^2$\\
$g_F$ & $L_0^r(-1 +{1\over N_F}-2N_F)-2L_1^r+L_2^r(-N_F-2N^2_F)+ L_3^r(-1+{1\over N_F}-N_F) $\\
      & $8N^2_F( L_4^r -2L_6^r)+ 4N_F( L_5^r -2L_8^r) $\\
$h_F$ & $-\frac{7}{96}+\frac{7}{32}\frac{1}{N_F^2}-\frac{7}{32}\frac{1}{N_F}
 +\frac{19}{256}N_F+\frac{1}{768}N_F^2$\\
\hline
 & 2-colour\\
\hline
$a_F$ & $\frac{1}{2}N_F$ \\
$b_F$ & $8 N_F L_4^r+4 L_5^r$\\
$c_F$ & $-\frac{1}{4}-\frac{3}{16}N_F-\frac{3}{16}N_F^2$\\
$d_F$ & $L_0^r(6 +{6\over N_F}-4N_F)+4L_1^r(1+2N_F-4N^2_F)-2L_2^r(4-N_F+2N^2_F)$
\\
      & $+ L_3^r(6+{6\over N_F}-10N_F)
         -4L_4^r( 1+N_F+N^2_F)-2 L_5^r(1+ {1\over N_F}+N_F)$\\
      & $+ 8N_F(2N_F L_6^r+L_8^r)$\\
$e_F$ & $\frac{7}{24}-\frac{1}{8N_F^2}-\frac{1}{8N_F}+\frac{29}{32}N_F-\frac{59}{96}N_F^2$\\
$f_F$ & $r^r_{FT}-8(2N_F L_4^r+L_5^r)^2$\\
$g_F$ & $L_0^r(1 +{1\over N_F}-2N_F)-2L_1^r+L_2^r(N_F-2N^2_F)+ L_3^r(1+{1\over N_F}-N_F) $\\
      & $8N^2_F( L_4^r -2L_6^r)+ 4N_F( L_5^r -2L_8^r) $\\
$h_F$ & $-\frac{7}{96}+\frac{7}{32}\frac{1}{N_F^2}+\frac{7}{32}\frac{1}{N_F}
 -\frac{19}{256}N_F+\frac{1}{768}N_F^2$\\
\hline
\end{tabular}
\end{center}
\caption{\label{tabdecay} The coefficients $a_F,\ldots,g_F$
appearing in the expansion of the decay constant.}
\end{table}
The subtractions needed for the QCD case have been derived in
general before in \cite{BCE2}. The adjoint and two-colour case can
be made finite by the following: 
\ba 
\Gamma^{(2)}_{FA} &=&
1-{3\over4} N_F + {1\over4}N_F^2 \,,
\nonumber\\
\Gamma^{(L)}_{FA} &=&-4 \Bigg[
\left(-3 +{3\over N_F} -2N_F\right)L^r_0
+2\left(1-2 N_F- 4N_F^2\right)L^r_1+\left(-4 -N_F- 2N_F^2\right)L^r_2
\nonumber\\
&&+\left(-3 +{3\over N_F}-5N_F\right)L^r_3
+{1\over N_F}(N_F-1)(2N_FL^r_4+ L^r_5) +8N_F^2L^r_6+4N_FL^r_8\Bigg]\,,
\nonumber\\
\Gamma^{(1)}_{FA} &=&- {1\over8}\left({1\over3}-{1\over N_F^2}+
{1\over N_F}-{53\over 12}N_F-{49\over 12}N_F^2\right)\,,
\nonumber\\
\Gamma^{(2)}_{FT} &=& 1+ {3\over4} N_F + {1\over4}N_F^2\,,
\nonumber\\
\Gamma^{(L)}_{FT} &=& -4 \Bigg[
\left(3 +{3\over N_F} -2N_F\right)L^r_0
+2\left(1 +2 N_F- 4N_F^2\right)L^r_1+\left(-4 +N_F- 2N_F^2\right)L^r_2
\nonumber\\
&&+\left(3 +{3\over N_F}-5N_F\right)L^r_3-{1\over
N_F}(1+N_F)(2N_FL^r_4 +L^r_5)+8N_F^2L^r_6+4N_FL^r_8\Bigg]\,,
\nonumber\\
\Gamma^{(1)}_{FT} &=&-{1\over8}\left({1\over3}-{1\over N_F^2}
-{1\over N_F}+{53\over 12}N_F-{49\over 12}N_F^2\right) \,. 
\ea
This
result agrees at NLO with \cite{GL4} for the QCD case and
\cite{Splittorff} for the 2-colour case. It also agrees with the
decay constant for two and three flavours in the QCD case as
calculated in \cite{Buergi,BCEGS,ABT1,GK}. The remaining results are
new.

The coefficient of the leading logarithm, $\overline A(M^2)^2$ is always
determined but note that the coefficient of the subleading logarithm for
the vacuum expectation value depends on LECs that can be determined from
the masses.

The expansions (\ref{expansionqbarq}), (\ref{expansionmass}) and
(\ref{expansiondecay}) have been written in terms of the lowest order
mass and decay constant. It is possible to reorder the series
in various ways. In particular one can rewrite
the series in terms of the physical masses and decay constants instead.
The logarithms come from physical particles propagating
so the form in terms of physical masses might be preferable.
There are some indications that in the case of two-flavour QCD this leads
to a better convergence, see e.g. \cite{CDlattice}. The physical
mass and decay constant expansion is referred to there as the $\xi$ expansion.
We thus rewrite (\ref{expansionqbarq}), (\ref{expansionmass}) and
(\ref{expansiondecay}) as
\be
O_\mathrm{phys} = O_\mathrm{LO}+ O_\mathrm{NLO}+ O_\mathrm{NNLO}\,,
\ee
with
\ba
O_\mathrm{NLO}
 &=& O_\mathrm{LO}
\left(\alpha_O \frac{\overline A(M^2_\mathrm{phys})}{F^2_\mathrm{phys}}+\beta_O\frac{M^2_\mathrm{phys}}{F^2_\mathrm{phys}}\right)\,,
\nonumber\\
O_\mathrm{NNLO}
 &=& O_\mathrm{LO}
\Bigg(\gamma_O \frac{\overline A(M^2_\mathrm{phys})^2}{F^4_\mathrm{phys}}
     +\frac{M^2_\mathrm{phys}\overline A(M^2_\mathrm{phys})}{F^4_\mathrm{phys}}
\left(\delta_O+\frac{\epsilon_O}{16\pi^2}\right)
\nonumber\\&&\qquad
     +\frac{M^4_\mathrm{phys}}{F^4_\mathrm{phys}}
  \left(\zeta_O+\frac{\eta_O}{16\pi^2}+\frac{\theta_O}{(16\pi^2)^2}\right)
\Bigg)\,.
\ea
We do this for $O=V,M,F$ for the vacuum-expectation-value, mass
and decay constant.
The coefficients in the two expansions are related by
\ba
\alpha_O &=& a_O\,,\qquad
\beta_O  = b_O\,,\qquad
\gamma_O = c_O+(2a_F-a_M)a_O\,,
\nonumber\\
\delta_O &=&d_O +(2b_F-b_M)a_O+(2a_F-a_M)b_O\,,\qquad
\epsilon_O = e_O + a_M a_O\,,
\nonumber\\
\zeta_O &=& f_O+(2b_F-b_M)b_O\,,\qquad
\eta_O = g_O + b_M a_O\,,\qquad \theta_O = h_O\,.
\ea
These can be easily evaluated using the results in Tabs.~\ref{tabqbarq}
to \ref{tabdecay}.

\section{Conclusions}
\label{conclusions}

In this work we have calculated the vacuum expectation value,
the meson mass and the meson decay constant
in effective field theory to NNLO for
the three cases with a simple underlying vector gauge groups
and $N_F$ equal mass fermions in the same representation.
We discussed the complex case (QCD), real representation (Adjoint)
and pseudo-real representation (2-colour).

The three flavour cases have been calculated earlier at NNLO
for the  QCD case for the mass, decay constant \cite{GK,ABT3}
and condensate \cite{ABT3}. For two flavour
QCD the NNLO expressions exists for the mass and decay constants
\cite{Buergi,BCEGS}.
For the $N_F$ flavour case the mass, decay constant
and the condensate can be found in
\cite{GL4} to NLO.
The NNLO expressions here are new. Note that the equal mass case
considered here leads to considerably simpler expressions than those
of \cite{GK,ABT1}.
We have a slightly different NLO divergence structure for the two-colour case
then \cite{Kogut} but agree with their explicit NLO expressions
for the mass, decay constant and vacuum expectation value.
Again the NNLO expressions here are new.
The adjoint case we have extended to NLO in general and to NNLO
for the mass, decay constant and vacuum expectation value.
Notice that for all three cases the coefficient of the leading logarithm at
NNLO is fully determined but that the coefficient of the subleading logarithm
at NNLO for the vacuum expectation value depends on LECs
that can be determined from the mass at NLO.

The main motivation behind this work is that these expressions
should be useful for extrapolations to zero mass
in lattice calculations for dynamical electroweak symmetry breaking.

\section*{Acknowledgements}

This work is supported by the
Marie Curie Early Stage Training program “HEP-EST” (contract number
MEST-CT-2005-019626),
European Commission RTN network,
Contract MRTN-CT-2006-035482  (FLAVIAnet),
European Community-Research Infrastructure
Integrating Activity
``Study of Strongly Interacting Matter'' (HadronPhysics2, Grant Agreement
n. 227431)
and the Swedish Research Council.

\appendix

\end{document}